\documentclass{elsart}

\usepackage{amsmath}
\usepackage{amsfonts}
\usepackage{latexsym}
\usepackage{graphicx}

\begin{document}

\runauthor{K.~Malarz and A.~Z.~Maksymowicz}
\begin{frontmatter}
\title{A Simple Solid-on-Solid Model\\ of Epitaxial Thin Films Growth:\\ Inhomogeneous Multilayered Sandwiches}
\author{K.~Malarz\thanksref{malarz}} and \author{A.~Z.~Maksymowicz}
\address{Department of Theoretical and Computational Physics,
Faculty of Physics and Nuclear Techniques,
University of Mining and Metallurgy (AGH)\\
al. Mickiewicza 30, PL-30059 Krak\'ow, Poland}
\thanks[malarz]{Corresponding author: e-mail: malarz@agh.edu.pl, fax: +48 12 6340010}
%% *******************************************************************
\begin{abstract}
In this work a simple solid-on-solid (SOS) model of inhomogeneous films epitaxial growth is presented.
The results of the computer simulations based on the random deposition (RD) enriched by a local relaxation which tends to maximize the number of a particle-particle lateral bonds (PPLB) against barriers blocking diffusion are presented.
The influence of strength of the atoms interactions and barriers for diffusion on film roughness and possible bridges between the layers in tri-layered sandwich system are considered.
For magnetic layers with a nonmagnetic spacer the bridging is responsible for direct magnetic coupling between layers.
Also variations of the spacer thickness, when scanning the film surface, is essential for uniformity of the magnetic coupling between the layers.
Number of bridges for very thin spacer, and variation of the spacer thickness are discussed in this work.
\end{abstract}
%% *******************************************************************

%% *******************************************************************
%% *******************************************************************
\end{frontmatter}

%% *******************************************************************
\section{Introduction} 
%% *******************************************************************
	In last decade the physics of growth thin solid films and ion/atoms behaviors on flat film surfaces became well understood and investigated due to rapid progress in experimental methods of preparing thin films (mainly MBE in UHV systems) and microscopy (mainly STM). 
Also many papers devoted to theory of growth of rough surfaces and interfaces and crystal were published (see \cite{gouyet91,herrmann86,meakin93,levi97} for review).

	Here, we present a simple SOS model~\cite{maksymowicz96,malarz99_1,malarz99_2} of epitaxial growth adopted for inhomogeneous $a/b/a$-like systems. 
The film growth described as a two step mechanism.
The first step is the randomly picked position of initial contact of the incident particle on the film surface.
This is followed by second step which is a relaxation process of local migration leading to final position of the particle sticking to the film.
We simplify the model to the limited surface diffusion and neglect migration to more distant sites then nearest neighbors (NN) only.
After the relaxation process particle sticks for the rest of the simulation.
Each particle is represented by a unit-volume cube which can occupy only discrete position in the lattice.
The simple cubic symmetry is assumed.
For a $L\times L$ substrate we deposit $\theta_a L^2$ particles of kind $a$, followed by $\theta_b L^2$ $b$-like particles and again $\theta_a L^2$ of $a$ particles.
The nominal thicknesses of $a$ and $b$ layers, are $\theta_a$ and $\theta_b$ respectively.
In relaxation process, each particle tends to maximize the number of PPLB.
This tendency is slowed down by the barrier $V$ for diffusion which decreases probability of atom movement.
The above growth model may be implemented as the following flowchart:
\begin{itemize}
\item for each newly arriving particle at random site $r=0$, calculate the number of atomic pairs $n^r_{aa}$, $n^r_{ab}$ and $n^r_{bb}$ at the place of the initial particle contact to the surface with all their four NN labeled by $r=1\ldots 4$,
\item calculate the particle total energies in all five positions $(r=0\ldots 4)$: $E^r=n^r_{aa}E_{aa}+n^r_{ab}E_{ab}+n^r_{bb}E_{bb}$, where $E_{ij}$ is the bonding energy between $i$- and $j$-kind atoms,
\item evaluate probabilities $p^r\propto \exp(-E^r)$, $r=0\ldots 4$, of picking out each of the five virtual final positions of the atom,
\item reduce the probability $p^r$ of movement into $r=1\ldots 4$ by factor $\exp(V_i)$, where $V_a$ and $V_b$ are the diffusion barriers for $a$- and $b$-kind of atoms, respectively,
\item pick out one of five proposed sites for the atom with probability given by $p^r$, $(r=0\ldots 4)$.
\end{itemize}
Values of $E$ and $V$ are expressed in $k_BT$ units, where $k_B$ is the Boltzmann constant, and $T$ denotes the absolute temperature.
Diffusion barrier $V$ is positive, while the negative $E$ are compatible with the assumed tendency of the system to maximize the number of PPLB.

%% *******************************************************************
\section{Results of simulation}
%% *******************************************************************
	The simulations were performed on $500\times 500$ large square lattice with periodic boundary conditions.
The nominal thickness $\theta_a$ of both $a$-layers was set up to ten monolayers (ML), while average thickness of $b$-spacer $\theta_b$ was varied from one to ten ML.
The cases $E_{aa}=E_{ab}=E_{bb}=0$, or $V_a=V_b\to+\infty$ which blocks any diffusion, yields the same results as RD model, with Poisson distribution of surface/interface heights.

%% ===================================================================
\subsection{Direct magnetic coupling}
%% ===================================================================
	Energy of magnetic coupling between magnetic layers separated by nonmagnetic spacer is often expressed by $E=-K\vec{M}_1\circ\vec{M}_2$ energy term, where $\vec{M}_1$ and $\vec{M}_2$ are magnetizations in magnetic layers.
The coupling $K$ depends strongly on nonmagnetic spacer thickness $\theta_b$ and may follow exponential law $K\propto\exp(-\theta_b/\theta_0)$, or it may have an oscillatory character against $\theta_b$ as it also is often observed.
The interaction energy $K(\theta_b)$ manifests itself experimentally by modifications of some of magnetic properties such as susceptibility or ferromagnetic resonance.
The coupling between magnetic layers separated by a nonmagnetic spacer was shown for example for NiFe/Cu/NiFe~\cite{maksymowicz91} and for Ni/Ag/NiFe samples~\cite{layadi90}.
In latter case for some samples also power decrease of $K$ with $\theta_b$ was observed.

\begin{figure}
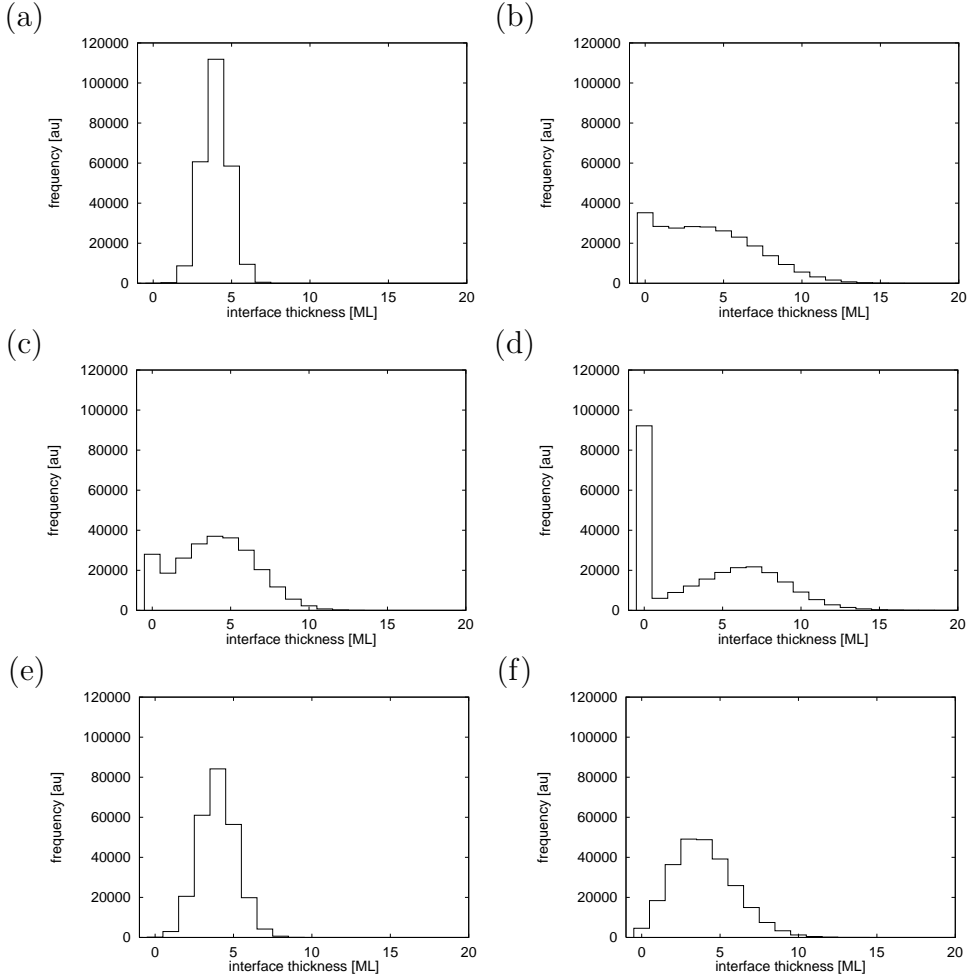

\centering
(a) \includegraphics[width=40mm,angle=-90]{aa-10_ab-10_bb-10_his.fig}
(b) \includegraphics[width=40mm,angle=-90]{aa000_ab-10_bb000_his.fig}\\
(c) \includegraphics[width=40mm,angle=-90]{aa000_ab000_bb-01_his.fig}
(d) \includegraphics[width=40mm,angle=-90]{aa000_ab000_bb-10_his.fig}\\
(e) \includegraphics[width=40mm,angle=-90]{aa-01_ab-10_bb-01_his.fig}
(f) \includegraphics[width=40mm,angle=-90]{aa000_ab000_bb000_his.fig}
\caption{Histogram of spacer heights for different model control parameters.
Nominal thickness was $\theta_b=4$~ML.
(a) $E_{aa}=-10$, $E_{ab}=-10$, $E_{bb}=-10$,
(b) $E_{aa}=0$, $E_{ab}=-10$, $E_{bb}=0$,
(c) $E_{aa}=0$, $E_{ab}=0$, $E_{bb}=-1$,
(d) $E_{aa}=0$, $E_{ab}=0$, $E_{bb}=-10$,
(e) $E_{aa}=-1$, $E_{ab}=-10$, $E_{bb}=-1$,
(f) $E_{aa}=0$, $E_{ab}=0$, $E_{bb}=0$.  }
\label{fig_his}
\end{figure}

	For simple RD model distribution of spacer thickness $h_b$ follows Poisson distribution $P(h;\theta)=\theta^h/h!\cdot\exp(-\theta)$, and probability of direct coupling between $a$-layers, corresponding to zero spacer height, decrease exponentially with average spacer thickness $\theta_b$: $P(0;\theta_b)=\exp(-\theta_b)$.
However, in film preparation technology, the growth conditions seldom correspond to RD model and so we expect deviations from Poisson distribution as presented in Fig.~\ref{fig_his}.
%% -------------------------------------------------------------------
%% -------------------------------------------------------------------
\begin{figure}
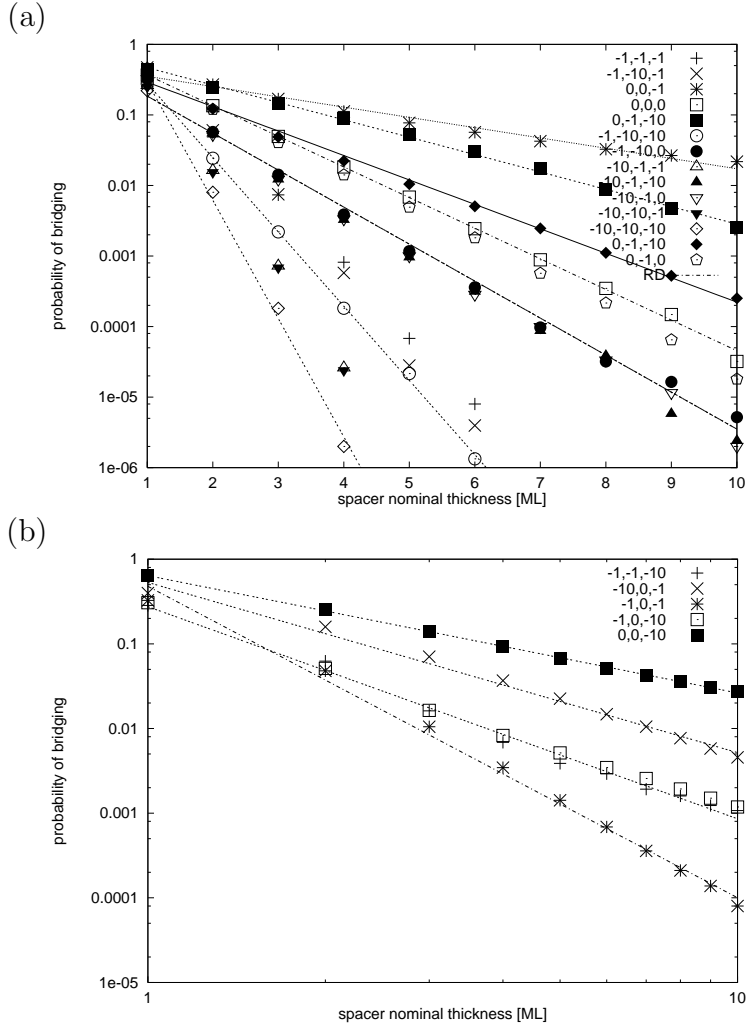

\centering
(a) \includegraphics[width=65mm,angle=-90]{pb-sem-log.fig}\\
(b) \includegraphics[width=65mm,angle=-90]{pb-log-log.fig}
\caption{Exponential (a) and power-law-like (b) dependence of the probability of bridging on the spacer thickness $\theta_b$ for different model control parameters $E_{aa}$, $E_{ab}$ and $E_{bb}$ shown in rigth upper corners.}
\label{fig_pb}
\end{figure}
%% -------------------------------------------------------------------
For different sets of model parameters for non-RD, we found that decrease of number of bridges (proportional to the direct coupling $K$) follow either exponential or the power law (Fig.~\ref{fig_pb}).

%% ===================================================================
\subsection{Spacer roughness}
\begin{figure}
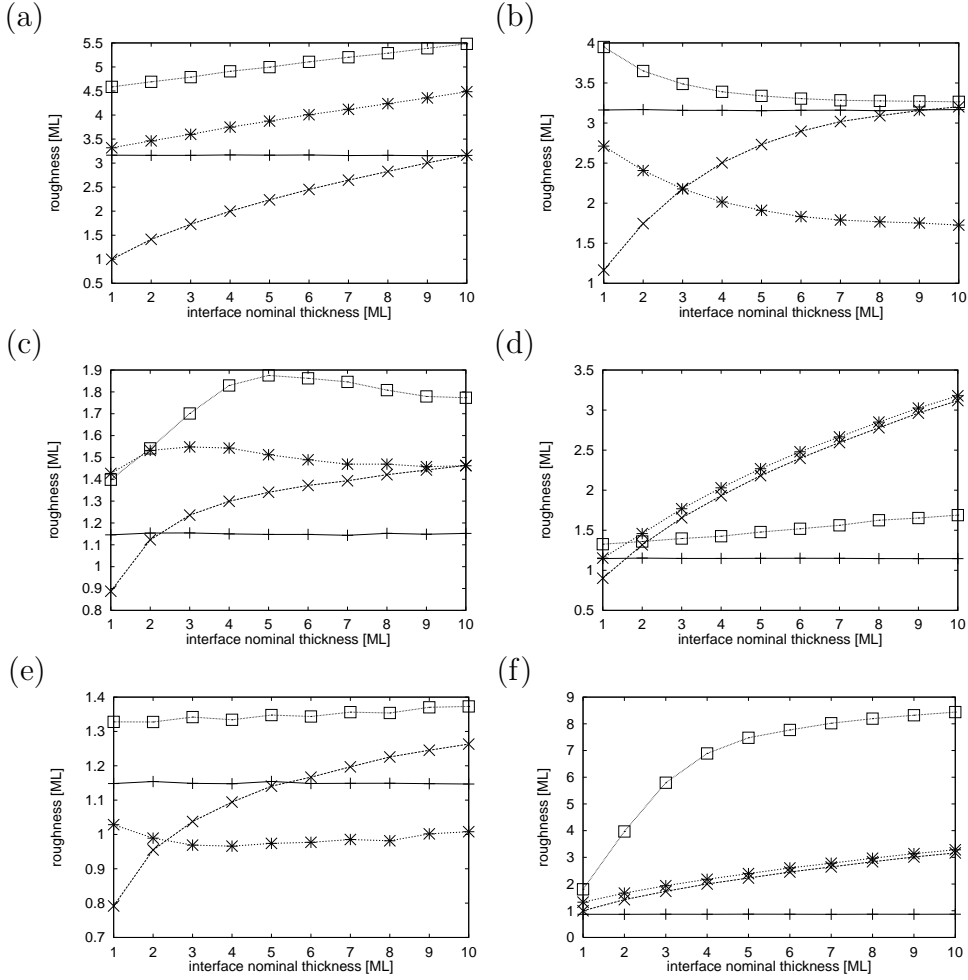

\centering
(a) \includegraphics[width=40mm,angle=-90]{aa000_ab000_bb000_sig.fig}
(b) \includegraphics[width=40mm,angle=-90]{aa000_ab-01_bb-10_sig.fig}\\
(c) \includegraphics[width=40mm,angle=-90]{aa-01_ab000_bb-01_sig.fig}
(d) \includegraphics[width=40mm,angle=-90]{aa-01_ab-10_bb000_sig.fig}\\
(e) \includegraphics[width=40mm,angle=-90]{aa-01_ab-10_bb-10_sig.fig}
(f) \includegraphics[width=40mm,angle=-90]{aa-10_ab000_bb000_sig.fig}
\caption{The lines of $+$, $\times$, $*$ and $\Box$ correspond to $\sigma_a$, $\sigma_b$, $\sigma_{ab}$, and $\sigma_{aba}$, respectively.
(a) $E_{aa}=0$, $E_{ab}=0$, $E_{bb}=0$ --- a RD  model,
(b) $E_{aa}=0$, $E_{ab}=-1$, $E_{bb}=-10$,
(c) $E_{aa}=-1$, $E_{ab}=0$, $E_{bb}=-1$,
(d) $E_{aa}=-1$, $E_{ab}=-10$, $E_{bb}=0$,
(e) $E_{aa}=-1$, $E_{ab}=-10$, $E_{bb}=-10$,
(f) $E_{aa}=-10$, $E_{ab}=0$, $E_{bb}=0$.}
\label{fig_sig}
\end{figure}

%% ===================================================================
	Variation of the spacer thickness is also essential for uniformity of the magnetic coupling between layers.
We use root-mean-square $\sigma$ of surface heights as a measure of surface roughness.
The dependence of $ab$-layer roughness on roughness of each of its components is described by: $\sigma_{ab}^2=\sigma_a^2+\sigma_b^2+2\langle h_ah_b\rangle-2\langle h_a\rangle \langle h_b\rangle$.
For RD, successive film heights are uncorrelated and thus: $\sigma_{ab}^2=\sigma_a^2+\sigma_b^2$ and $\sigma_i^2=\theta_i$ for $i=a,~b,~ab$.
The situation for non-RD case is more complex (see Fig.~\ref{fig_sig}b-f).
%% -------------------------------------------------------------------
%% -------------------------------------------------------------------
	We would like to consider roughness $\sigma_b$ dependence on the spacer thickness $\theta_b$ for different model control parameters.
For homogeneous films growth models roughness $\sigma$ scaling with film thickness with Family-Vicsek law~\cite{family85}: $\sigma\propto L^\alpha f(\theta/L^z)$ with $f(x\to 0)\propto x^\beta$ and $f(x\to\infty)\propto 1$.
For large enough lattice size, and not too large average film thickness, this dependence is given by a power law: $\sigma\propto\theta^\beta$.
The dependence of exponent $\beta$ on model control parameters was discussed in details in Ref.~\cite{malarz99_1}.  
This way we are able to predict roughness $\sigma_a(\theta_a)$ of the first $a$ layer.
However, determination of $\sigma_b$ is more difficult.
Firstly, $\theta_b$ is small to guarantee clear dependence $\sigma_b$ on $\theta_b$.
Secondly, the growth of $b$ spacer is also controlled by $E_{ab}$ until $b$-like coverage is sufficient to creating $bb$-like PPLB when becomes dependent on $E_{bb}$.
Fig.~\ref{fig_sig} shows that $\sigma_b$ increases with increasing of $\theta_b$ independently on $E$ and $V$ parameters.
Deviations from basic power scaling law, however, may be observed.

%% ===================================================================
\section{Conclusion}
%% ===================================================================
	We found from computer simulations that distribution of spacer heights around average thickness changes from Poisson distribution (or bell-shaped for larger thickness), to peak-shaped with a decrease of diffusion barriers $V$ and/or increasing $E$, for a tendency of particles to create more PPLB.
The latter case helps to produce more uniform of magnetic coupling between layers in the sandwich tri-layer systems.

	Energy of magnetic coupling between two magnetic moments $\vec{\mu}_1$ and $\vec{\mu}_2$ may be expressed by Heisenberg term $E=-K \vec{\mu}_1\circ\vec{\mu}_2$, where $K$ is the coupling constant.
It is common to assume that this interaction is a short range one.
For direct exchange interaction $J$, $K\propto J$ and $J\ne 0$ only if $h_b=0$.
The direct coupling means that $h_b=0$ and strength of coupling $K$ between magnetic layers with magnetizations $\vec{M}_1=\sum \vec{\mu}_1$ and $\vec{M}_2=\sum\vec{\mu}_2$ may by evaluated as $K=NJ$, where $N=P(0;\theta_b)\cdot L^2$ is the number of bridges between magnetic layers.
Usually, the decrease of $K$ with increasing spacer thickness follows exponential or power law.
We found from computer simulations that number of bridges responsible for direct coupling is compatible with the above predictions, and either exponential or power law may be obtained for specific set of model parameters.

%% ===================================================================
\begin{ack}
%% ===================================================================
This work and machine time in ACC-CYFRONET-AGH is financed by Polish Committee for Scientific Research \mbox{(KBN)} with grants no. 8~T11F~02616 and KBN/\-S2000/\-AGH/\-069/\-1998, respectively.
\end{ack}

%% ===================================================================

\end{document}